\documentclass[a4paper,fleqn]{cas-dc}

\usepackage[comma,square,numbers,sort&compress]{natbib}
\usepackage{amsmath,amssymb,amsfonts}
\usepackage{algorithmic}
\usepackage{graphicx}
\usepackage{textcomp}
\usepackage{booktabs}
\usepackage{siunitx}

\usepackage{multirow}

\begin{document}
\let\WriteBookmarks\relax
\def\floatpagepagefraction{1}
\def\textpagefraction{.001}

\shorttitle{GPU propagation and visualisation of particle collisions with ALICE magnetic field model}

\shortauthors{Piotr Nowakowski et~al.}


\title [mode = title]{GPU propagation and visualisation of particle collisions with accurate model of ALICE detector magnetic field}

\address[1]{Faculty of Electronics and Information Technology, Warsaw University of Technology, Nowowiejska 15/19, 00-665 Warsaw, Poland}

\author[1]{Piotr Nowakowski}[orcid=0000-0001-8971-0874]
\ead{piotr.nowakowski.dokt@pw.edu.pl}
\cormark[1]

\author[1]{Przemys\l{}aw Rokita}[orcid=0000-0002-4433-2133]

\ead{P.Rokita@ii.pw.edu.pl}

\address[2]{Faculty of Physics, Warsaw University of Technology, Koszykowa 75, 00-662 Warsaw, Poland}

\author[2]{\L{}ukasz Graczykowski}[orcid=0000-0002-4442-5727]
\ead{lukasz.graczykowski@pw.edu.pl}

\cortext[cor1]{Corresponding author}

\begin{abstract}
The ALICE Collaboration at CERN developed a 3D visualisation tool capable of displaying a representation of collected collision data (particle trajectories, clusters and calorimeter towers) called the Event Display. The Event Display is constantly running in the ALICE Run Control Center as part of the Quality Assurance system, providing the monitoring personnel with visual cues about possible problems of both hardware and software components during periods of data gathering. In the software, particle trajectories (which are curved due to the presence of a magnetic field inside the detector) are generated from the physical parameters of detected particles, such as electrical charge and momentum. Previously this process in the Event Display used a uniform, constant magnetic field for these calculations, which differs from the spatial variations of the real magnetic field and does not model one of the two magnets used in the detector. Recently, a detailed model of the ALICE magnetic field was made available as a shader program for execution on the GPU. In this work we attempt to implement the reconstruction algorithm in a shader form as well, allowing us to combine it with the detailed model to create a full solution for rendering trajectories from collision event data directly on the GPU. This approach has several possible advantages, such as better performance and the ability to alter the magnetic field properties in real time. This was not previously done for ALICE and as such could be used in the future to upgrade the Event Display.

\end{abstract}


\maketitle

\section{Introduction}

\textbf{ALICE} (\textbf{A} \textbf{L}arge \textbf{I}on \textbf{C}ollider \textbf{E}xperiment \cite{alicedetector}) is one of the experiments located at the \textbf{L}arge \textbf{H}adron \textbf{C}ollider --- \textbf{LHC}~\cite{lhcmachine}. The data ALICE detector gathers can be visualised in a software application called the Event Display~\cite{alieve}. The Event Display can show both the particle trajectories (tracks) and towers (3D histograms of energy measured by calorimeters). It is an integral  part of a sophisticated Quality Assurance system that monitors the detector status and the correct functioning of the online data processing done on its computer cluster (see Section~\ref{sec:alice}). The online processing performs data compression and produces data for calibration from signals received from the detector. Properties of charged particles, such as momentum, can be calculated from their curvature in magnetic fields. ALICE uses two magnets for this purpose: a solenoid magnet in the central detector and a dipole magnet in the forward spectrometer (see Section~\ref{sec:alice}).

Particle trajectories in the Event Display are generated (propagated) from reconstructed parameters. Previously for this purpose, a simplified uniform magnetic field model was used in the track visualisation software, which differs (by around 3\%) from the spatial variations of the real magnetic field of the solenoid and, more importantly, does not model the dipole field~\cite{nowakowski}.

A more accurate magnetic field representation based on actual field measurements is available in the ALICE software package. In our previous publication~\cite{nowakowski} we have studied how the accurate model could be implemented as a shader program for execution on a GPU, allowing its use in interactive magnetic field visualisations.
In this work, we explore the possibility of implementing the particle propagation code in a similar way (in a shader). It gives us the ability to perform this task in real-time, as well as allows us to benefit from the accurate magnetic field model for visualisation of tracks with improved fidelity. The full source code of the propagator as well as of the benchmark software used for tests is available in a GitHub repository~\footnote{Developed software repository: \url{https://github.com/pnwkw/gpu\_propagator}}.

The particle propagation algorithm used presently in visualisation originates from ROOT, a framework for data analysis designed for high energy physics \cite{root}. Besides the mentioned uniform magnetic field model, this is a CPU implementation that performs the propagation in an iterative fashion and as such requires uploading the generated vertices to the GPU to be then displayed with OpenGL~\cite{openglbible}, DirectX~\cite{directx} or Metal~\cite{metal}.

For the upcoming period of data-taking at the LHC after a long break (during which ALICE has upgraded several detector components) a new software framework was developed for ALICE called Online-Offline, or \(\rm O^2 \)~\cite{otwodesign, otwotechnical}. One of the core features of the framework is the ability to write architecture-agnostic code which can be compiled to single-threaded or multi-threaded (with OpenMP) programs for the CPU as well as to CUDA / OpenCL / HIP programs for NVIDIA and AMD GPUs. Due to this the particle propagator for visualisation in \(\rm O^2 \) can be run on both CPU and GPU. Still, the GPU frameworks mentioned above can not display 3D graphics and as such require additional code for the actual rendering, e.g. in OpenGL, just like it is the case for the ROOT propagator.

The majority of modern graphics is generated from models created by artists, which are later arranged by map designers to form rich 3D worlds. Models are usually only slightly transformed in the graphics pipeline during rendering, e.g. via skeletal rigging or tessellation. Although not as commonly used, procedural generation of models is possible directly on the GPU using a programmable stage of the pipeline called the geometry shader. In literature, the geometry shader is utilized in vastly different scenarios, for example in a power grid CAD to display power line networks~\cite{geometry_power}, to perform scene voxelization for use in complex lightning effects~\cite{geometry_voxel} and to create a particle system for displaying fire, smoke, wind and magic effects in a video game~\cite{geometry_particles}, among others. In our opinion, the single most similar work to what we have achieved, based on the type of primitive rendered, was done for fiber tracking visualisation of white matter in the brain~\cite{geometry_brain}.

The rest of this paper is structured as follows. In Section \ref{sec:alice} the ALICE detector is briefly introduced. In Section \ref{sec:prop} an overview of the original particle propagator algorithm is presented. Section \ref{sec:impl} describes our OpenGL implementation in detail, which is followed by an explanation of how it was tested in Section \ref{sec:methodology}. Section \ref{sec:res} presents the experimental results. Finally, Section \ref{sec:conc} contains the conclusion drawn from the results, as well as ideas for further research.

\section{ALICE detector}
\label{sec:alice}

\begin{figure}[ht]
    \centering
    \includegraphics[width=\linewidth]{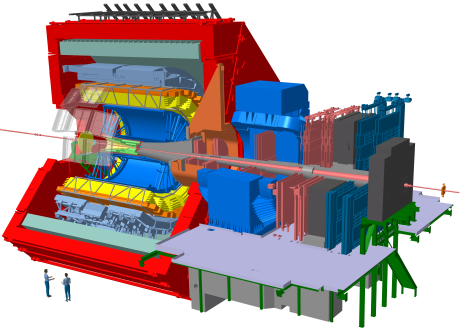}
    \caption{The schematics of the ALICE detector~\cite{alicedetector}.}
    \label{fig:alice}
\end{figure}

The detector is located \SI{56}{\meter} underground in a concrete cavern  at one of the LHC beam crossing sections located on the territory of Sergy in France. Most of the scientific apparatus is located inside a large solenoid magnet (see Figure \ref{fig:alice}, red element in the picture). In the center of this cylindrical magnet the point of collision (called the interaction point) is located. Different components measuring different particle properties are stacked in layers around the interaction point, through which particles created in the collision event pass sequentially. Pixel detectors are installed closest to the center; followed by detectors for particle identification: Time of Flight detector (TOF) and 
Transition Radiation Detector (TRD); and finally a photon spectrometer and electromagnetic and hadronic calorimeters. The tracking system records the trajectories of particles, while calorimeters stop them and measure their energy. Physicists can later identify what was created during a collision based on the information gathered.

To detect muons a special section in front of the barrel, called the ``Muon Arm'', was built. The connection between these two elements is laid out with a thick absorber material, which stops the majority of particles other than the muons. The Muon Arm contains a dedicated spectrometer and has its own dipole magnet for bending trajectories of muons, producing an even stronger field than the main solenoid (solenoid --- \SI{0.5}{\tesla}, dipole --- \SI{0.7}{\tesla}).

The ALICE experiment, during data taking periods, produces much more data than is feasible for permanent storage. For this reason the data processing is handled by two types of computer clusters, \textbf{FLP} --- \textbf{F}irst \textbf{L}evel \textbf{P}rocessors and \textbf{EPN} --- \textbf{E}vent \textbf{P}rocessing \textbf{N}odes~\cite{epn_fpl}. The FLP cluster (located underground, near the detector itself) handles the raw electronic measurements performed by the instruments and reduces the incoming bandwidth from \SI{3.4}{\tebi\byte/\second} to \SI{500}{\gibi\byte/\second} by zero suppression. This stream is received by the EPN cluster on the ground above the detector, which performs the actual reconstruction. The full reconstruction process applies reduction and compression to the data which is then written to local storage at a rate of \SI{100}{\gibi\byte/\second}. The goal of the reconstruction process is to calculate the physical properties of detected particles, such as initial seen position, momentum, and electrical charge. A software tool called Event Display can be used to browse through the stored files and graphically inspect collision results. Particle trajectories are recreated for visualisation by the tool itself based on the available parameters.

\section{Particle Propagator in ROOT framework}
\label{sec:prop}

To generate a track, the particle propagator requires a particle's charge \(q\), a current position vector \(\vec{V}\), a magnetic field vector at the current position \(\vec{B}\) and a momentum vector \(\vec{P}\). To calculate the next step position of the particle in a helical path, a local coordinate system (with base vectors: parallel to the field \(\vec{e_1}\) and two vectors defining a plane perpendicular to the field, \(\vec{e_2}\) and \(\vec{e_3}\)) need to be obtained first.

The parallel vector \(\vec{e_1}\) is equal to the normalized \(\vec{B}\) vector. To obtain \(\vec{e_2}\) and \(\vec{e_3}\), the momentum vector \(\vec{P}\) has to be first split into parallel \(\vec{P_{\|}}\) and transversal \(\vec{P_{\perp}}\) components.

\(\vec{P_{\|}}\) can be obtained by performing the dot product of \(\vec{P}\) and \(\vec{e_1}\), which calculates the magnitude of parallel component. Multiplication of \(\vec{e_1}\) by this magnitude results in \(\vec{P_{\|}}\). The transversal component of the momentum is a result of subtracting \(\vec{P_{\|}}\) from \(\vec{P}\). The transversal vector, \(\vec{e_2}\) is obtained by normalization of \(\vec{P_{\perp}}\).

With \(\vec{e_1}\) and \(\vec{e_2}\), the final vector required for helix, \(\vec{e_3}\), can be calculated by performing a cross product between them. If \(q\) is negative, the direction of \(\vec{e_3}\) has to be then reversed.

The particle propagator performs the following equations to obtain \(\vec{e_1}\), \(\vec{e_2}\), \(\vec{e_3}\) base vectors:

\begin{align}
    \vec{e_1} &= normalize(\vec{B}), \\
    \vec{P_{\|}} &= (\vec{P} \cdot \vec{e_1}) * \vec{e_1}, \\
    \vec{P_{\perp}} &= \vec{P} - \vec{P_{\|}}, \\
    \vec{e_2} &= normalize(\vec{P_{\perp}}), \\ 
    \vec{e_3} &= sign(q) * (\vec{e_1} \times \vec{e_2}).
\end{align}
\noindent Please note that the order of vectors in the cross product is non-standard (compared to the textbook equation for Lorentz force, \( \vec{F} = q \vec{v} \times \vec{B} \)) because the propagator uses inverted magnetic field direction convention~\footnote{Code reference: \url{https://github.com/root-project/root/blob/master/graf3d/ eve/src/TEveTrackPropagator.cxx\#L55}}.

\begin{figure}[ht]
    \centering
    \includegraphics[width=\linewidth]{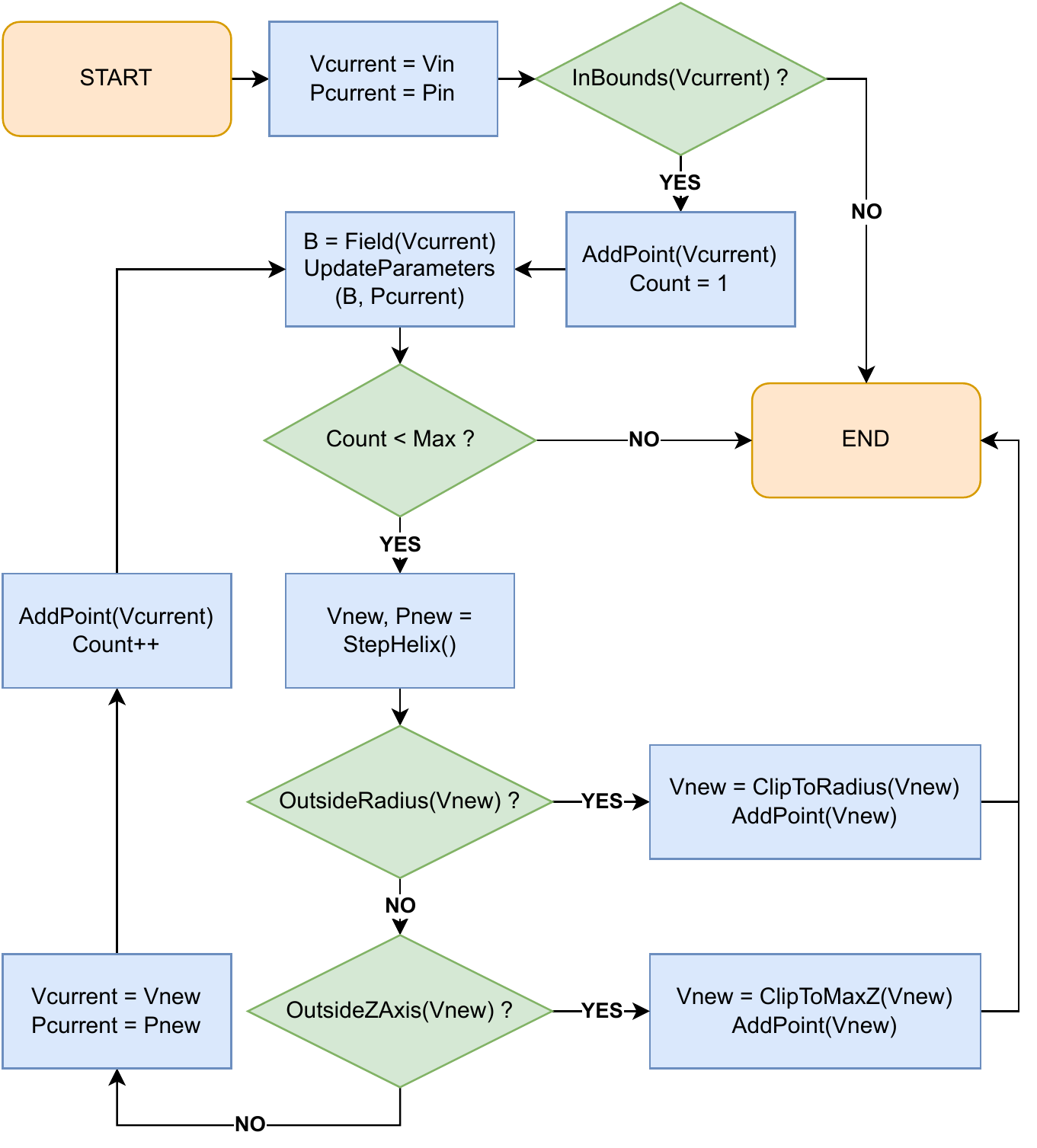}
    \caption{ROOT particle propagator algorithm diagram. Function \texttt{Update Parameters()} performs the base vector calculations for helical motion, needed by \texttt{StepHelix()}. Functions \texttt{ClipTo*()} perform boundary clipping. See Section \ref{sec:prop} for detailed description of the algorithm.} 
    \label{fig:propagator_algo}
\end{figure}

The updated position and momentum can be calculated using the parametric equation for the helical motion:
\begin{align}
    R & = \frac{\left| \vec{P_{\perp}} \right|}{B2C * \left| \vec{B} \right| * \left| q \right|}, \\
    \vec{V_{new}} & = \vec{V_n} + R \frac{\left| P_{\|} \right|}{\left| P_{\perp} \right|} \varphi * \vec{e_1} \\
    & + R \sin \varphi * \vec{e_2} \\
    & + (R * (1 - \cos \varphi )) * \vec{e_3}, \\[1em]
    \vec{P_{new}} &= \vec{P_{\|}} + \left| \vec{P_{\perp}} \right| \cos \varphi * \vec{e_2}, \\
    & + \left| \vec{P_{\perp}} \right| \sin \varphi * \vec{e_3},
\end{align}
\noindent where \( B2C \) is a momentum to curvature conversion constant\footnote{\SI[per-mode=symbol]{0.299792458e-2}{\giga\eV\per{\centi\metre\tesla\clight\elementarycharge}}}, \( R \) is the helix radius and the azimuthal angle \( \varphi \) is the adjustable step size. These two steps (updating the base vectors and calculating new positions) are repeated in a loop until propagation bounds (configured by the user as maximum distance from the center) or the maximum number of steps (configured by the user directly) are reached. The general overview of the algorithm is available as a diagram in Figure \ref{fig:propagator_algo}.

\begin{figure}[ht]
    \centering
    \includegraphics[width=0.8\linewidth]{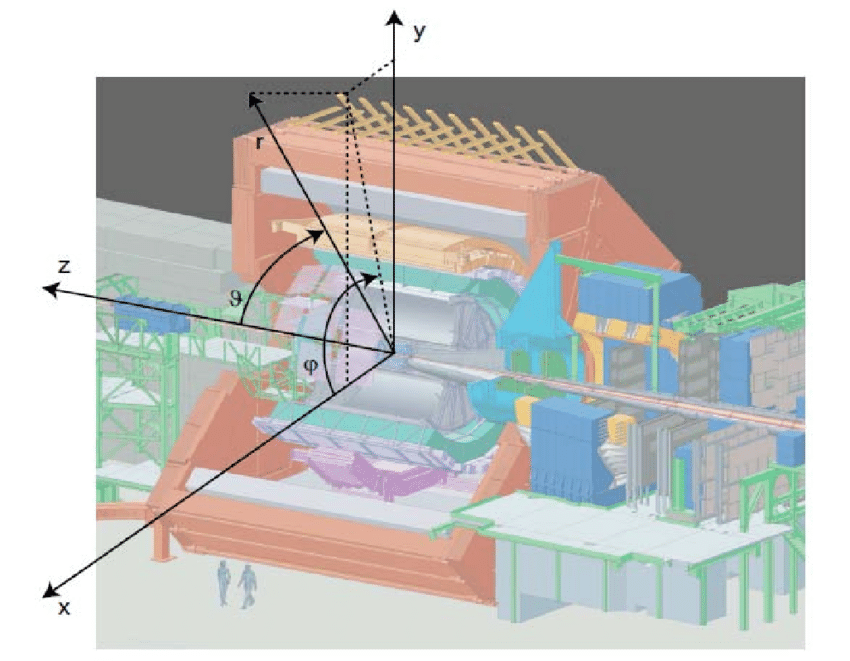}
    \caption{The ALICE detector (central barrel and the muon spectrometer) and its coordinate system \cite{Cavicchioli:2011kea}.}
    \label{fig:coordsystem}
\end{figure}

Propagation bounds are defined as a maximum distance along the LHC beam axis (in the ALICE coordinate system --- \( Z \) axis, see Figure~\ref{fig:coordsystem}) and a maximum radius on the plane perpendicular to the beam axis (in the ALICE coordinate system --- \( XY \) plane, see Figure~\ref{fig:coordsystem}). If the radius boundary is reached or exceeded by the point from the next step, its position is clipped to it. The clipping is implemented using a linear interpolation between the previous and next step, with the weight adjusted so that the resulting point lands on the boundary exactly. A similar operation is done for the \( Z \) axis boundary --- here the linear interpolation puts the new point at the maximum \( Z \) distance.

\section{Implementation}
\label{sec:impl}

Based on the ROOT propagator implementation (contained in the \texttt{TEveTrackPropagator} class)~\cite{root} (described in the Section above) we have created a version suitable for execution on a GPU in the GLSL shader language. The best fit for the propagator is the geometry shader stage of the OpenGL rendering pipeline~\cite{openglbible}, due to the generative nature of performed operations (an array of points on the particle trajectory is created from a single starting point).

The structure of the geometry shader is very similar to the original algorithm (the main difference being utilization of GLSL built-in vector operators --- e.g. scalar multiplication, cross and dot products --- in the shader versus ``manual'' implementation of them in the original code), except for the output --- generated vertices are not added to a buffer, but instead multiplied by a Model-View-Projection matrix and sent to the rasterizer by \texttt{EmitVertex()} call. The shader can be configured to output \texttt{points} (used for benchmarking purposes --- see Section \ref{sec:methodology}) or \texttt{line\_strip} for displaying tracks (see Figure \ref{fig:visual_inspection}).

The detailed magnetic field model implementation~\cite{nowakowski} offers a couple of options with different (accuracy, resource usage, performance) trade-offs, which can be generally distinguished by the method of magnetic field data storage:

\begin{itemize}
    \item Shader Storage Buffer (SSBO) --- \texttt{ssbo},
    \item 3D Texture --- \texttt{texture},
    \item Sparse 3D Texture --- \texttt{sparse},
    \item Chebyshev (reimplementation of the original model in a shader)  --- \texttt{glsl},
    \item Constant field (uniform, no boundaries)  --- \texttt{const},
    \item Constant Barrel field (const field limited to central barrel volume) --- \texttt{const\_barrel}.
\end{itemize}
\noindent Shader Storage Buffer method uses a simple array to store field samples. The 3D Texture method uses a floating point texture to store field samples. Sparse Texture optimizes the memory usage of Texture by not keeping parts of no field in memory. Chebyshev's method implements the original ALICE magnetic field model algorithm.

We have tested our propagator implementation with every option the detailed model offers.

\section{Experimental Methodology}
\label{sec:methodology}

As a source of data to run both the original and our GPU implementation of the propagator for benchmark purposes we have used particle data taken from simulated collisions generated by the ALICE Monte Carlo software.

Tests were performed with default propagator settings. Each test was repeated 10 times to reduce the effect of unpredictable variables such as CPU load --- results shown in Section \ref{sec:res} are averages.

\subsection{Retrieving data from the GPU}

To evaluate the differences between tracks generated by our geometry shader and the original ROOT implementation, we need to fetch the calculated vertices to the CPU side. To do this we have used the \texttt{Transform Feedback} feature, which instructs the GPU to write primitives (prepared for rasterization) also to a secondary buffer. The contents of this secondary buffer can be copied to CPU memory after the draw call is finished.

To measure time spent in the geometry shader as accurately as possible, we decided to disable pixel output (via \texttt{glEnable(GL\_RASTERIZER\_DISCARD)} function call) to skip clipping, rasterization, and fragment processor stages~\cite{openglbible}.

\subsection{Execution time measurement}

We measure render time as time spent between issuing the \texttt{glDrawArrays} command and copying the generating vertices using \texttt{glGetNamedBufferSubData} with the highest precision clock available in the standard library, \texttt{std::chrono::high\_resolution\_clock}.

Under normal circumstances, measuring the time elapsed in draw functions such as \texttt{glDrawArrays} does not result in an accurate performance metric, because the graphics driver does not wait for GPU to finish --- it simply adds the requested drawing command to the rendering queue. In our case however, the buffer copy mentioned above introduces an explicit synchronization point in the GPU processing queue, i.e. the driver will block the code execution on \texttt{glGetNamedBufferSubData} until the relevant operations are finished on the GPU. This allows us to accurately measure the execution time of our shader. On the other hand, this causes the rendering pipeline to be fully emptied, which does not happen in normal circumstances. Pipeline processing partially compensates for the time needed to render each image frame --- due to this reason the obtained results should be treated as the worst possible case. 

\subsection{Accuracy measurement}

To quantify the difference between tracks generated by the ROOT and our GLSL implementation we have used the Root Mean Square Error which is calculated as the difference between positions at each step in propagation. RMSE is calculated on each axis (\(X,Y,Z\)) individually.

As mentioned in Section \ref{sec:prop}, the algorithm generates track points until the configured count limit is reached or when it reaches boundaries, in which case it is cut short. The variable length of each track is problematic for interpreting results from the GPU, as all generated vertices for all tracks are tightly packed in the fetched buffer. This unfortunately makes start and end indices for each track unknown. To address this issue we have added an option to the shader to always pad the track with zero vectors up to the configured max length. With this feature enabled, each track starts at regular intervals in the buffer. Due to additional work required by the geometry shader when padding is used, we have enabled this option only for measurements of accuracy and kept it disabled for measurements of execution time described above.

\section{Results}
\label{sec:res}

Tests were performed on the following hardware:

\begin{itemize}
    \item Desktop computer --- NVIDIA GeForce RTX 2080 Ti (\SI{11}{\gibi\byte} Video RAM), AMD Ryzen Threadripper 1920X \SI{3.5}{\giga\hertz},
    \item ThinkPad X1 Extreme laptop --- NVIDIA GeForce GTX 1050 Ti with Max-Q Design (\SI{4}{\gibi\byte} Video RAM), Intel Core i7-8850H \SI{2.6}{\giga\hertz}.
\end{itemize}
\noindent On the laptop computer, the benchmark was run on Windows 10 (with NVIDIA Graphics Driver 460.39 installed). On the desktop computer, the benchmark was run on Ubuntu 20.04.4 LTS with NVIDIA Graphics Driver 470.103. On the Windows machine the program was compiled with \textit{msvc} version 19.21 and \texttt{/O2 /Ob2} optimization flags. On the Ubuntu machine the program was compiled with \texttt{gcc} version 10.1.0 and \texttt{-O3} optimization flags.

\begin{figure}[ht]
    \centering
    \includegraphics[width=\linewidth]{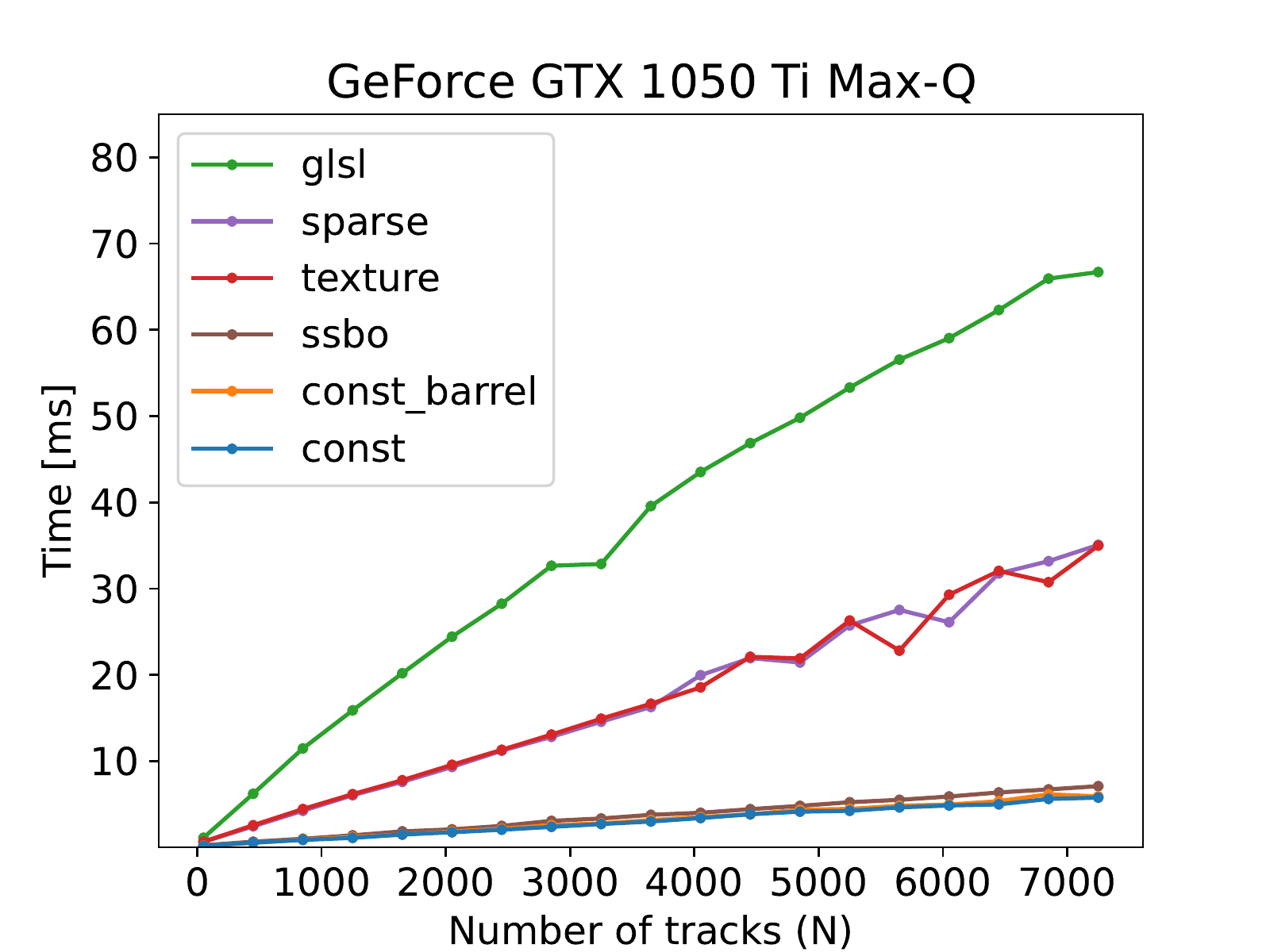}
    \caption{Average frame render time in milliseconds for track count from 50 to 7250 on the GeForce GTX 1050 Ti Max-Q, for every implementation of the magnetic field data (see Section~\ref{sec:impl}).}
    \label{fig:time_1050ti}
\end{figure}

\begin{figure}[ht]
    \centering
    \includegraphics[width=\linewidth]{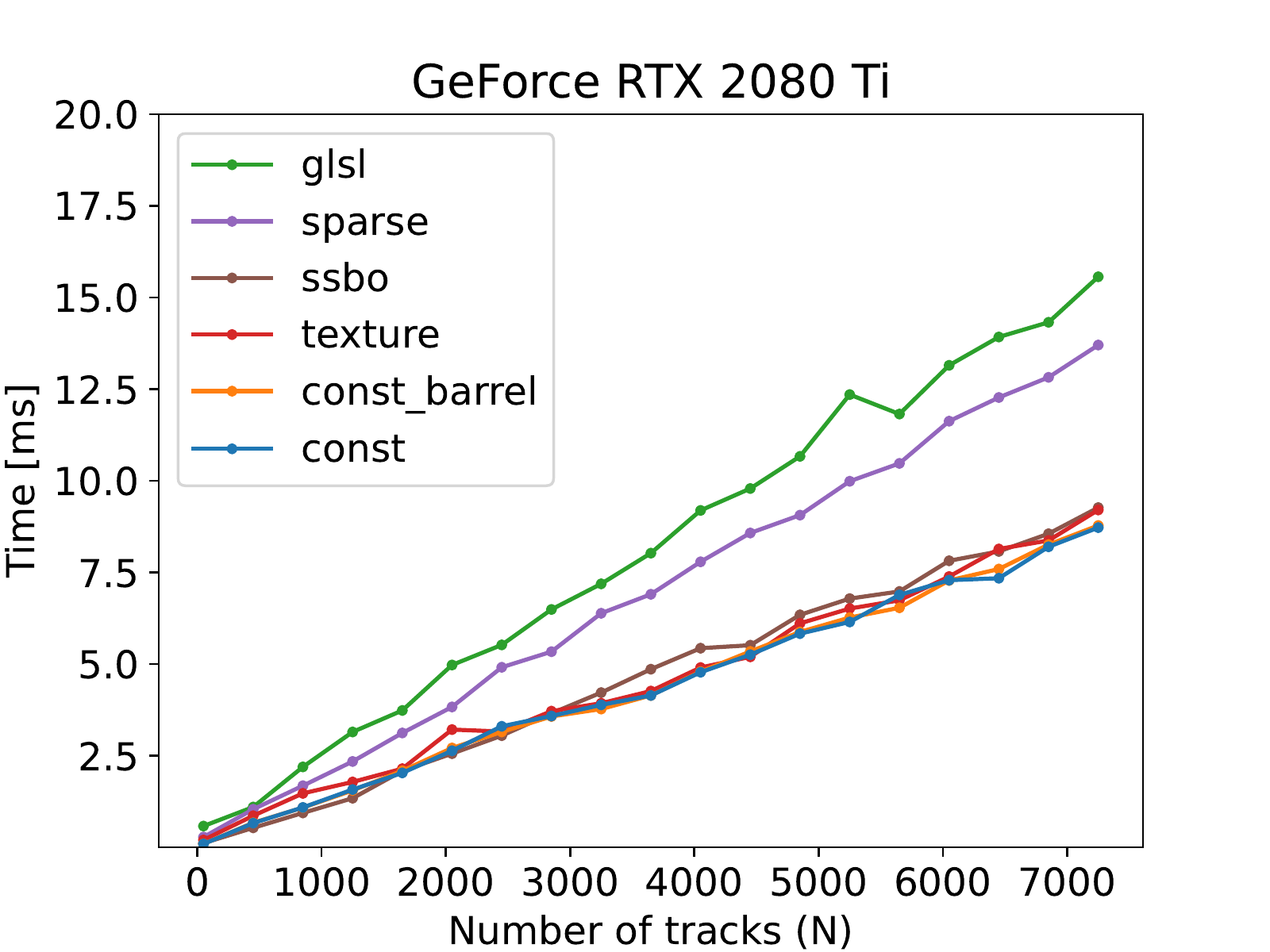}
    \caption{Average frame render time in milliseconds for track count from 50 to 7250 on the GeForce RTX 2080, for every implementation of the magnetic field data (see Section~\ref{sec:impl}).}
    \label{fig:time_2080}
\end{figure}

Figures \ref{fig:time_1050ti} and \ref{fig:time_2080} show the average render time for the number of tracks from 50 to 7250. For reference, a single Pb--Pb collision (after removal of noise and low-momentum tracks, which are not interesting from a physics perspective) generates on average 4000 entries. It can be seen that results from the older tested card (GTX 1050 Ti) follow a similar trend to what was measured in \cite{nowakowski}, i.e. the constant field methods and the shader storage buffer object method are the fastest (\SIrange[range-phrase=--, range-units=single]{7}{10}{\milli\second} at 7250 tracks, which corresponds to 100--143 frames per second), followed by texture-based methods (\SIrange[range-phrase=--, range-units=single]{25}{28}{\milli\second} at 7250 tracks, which corresponds to 35--40 frames per second) and finally the Chebyshev polynomial method (\SI{67}{\milli\second} at 7250 tracks, which corresponds to 15 frames per second). At the mentioned typical amount of tracks (4000) the constant methods produce 250 FPS, the buffer method produces 175 FPS, the texture methods produce 50 FPS, and the Chebyshev method produces 19 FPS.

On the 2080 Ti the constant field methods and the shader storage buffer object method performed with \SI{9}{\milli\second} at 7250 tracks, which corresponds to 111 frames per second. This is very similar to the level of results on 1050 Ti, which suggests that in this particular case we hit some kind of a bottleneck other than the power of the card itself, most likely the penalty of the explicit GPU synchronization (caused by buffer readback) used in the benchmark method. Here, the plain texture method was as performant as the methods just mentioned. The processing-intensive Chebyshev method in this case performed much faster, at the same level as the sparse texture method (\SIrange[range-phrase=--, range-units=single]{13}{16}{\milli\second} at 7250 tracks, which corresponds to 62--77 frames per second). At 4000 tracks, the constant, plain texture, and shader storage buffer methods achieved 208 frames per second. The sparse texture method achieved 128 frames per second, while the Chebyshev method achieved 109 frames per second.

Table \ref{tab:rmse} presents the average RMSE per track, calculated for the highest (7250) tested amount of tracks. It can be seen that our propagator used with a constant field (which mirrors the ROOT propagator exactly) results in a very low, close to zero error. This proves that our implementation is working correctly. The error is not exactly zero most likely due to a combination of rearrangement of floating point operations generated by the shader compiler (resulting in a different set of rounding errors) as well as varying level of IEEE-754 (floating point) standard support on the GPUs \cite{floataccuracy}. The same error level was obtained with the \textit{Const Barrel} method. The rest of the tested methods perform with a higher --- but similar to each other --- level, achieving an average deviation of \SI{5}{\centi\meter} per track when compared to the ROOT propagator with a constant field.

\begin{figure}[ht]
    \centering
    \includegraphics[width=\linewidth]{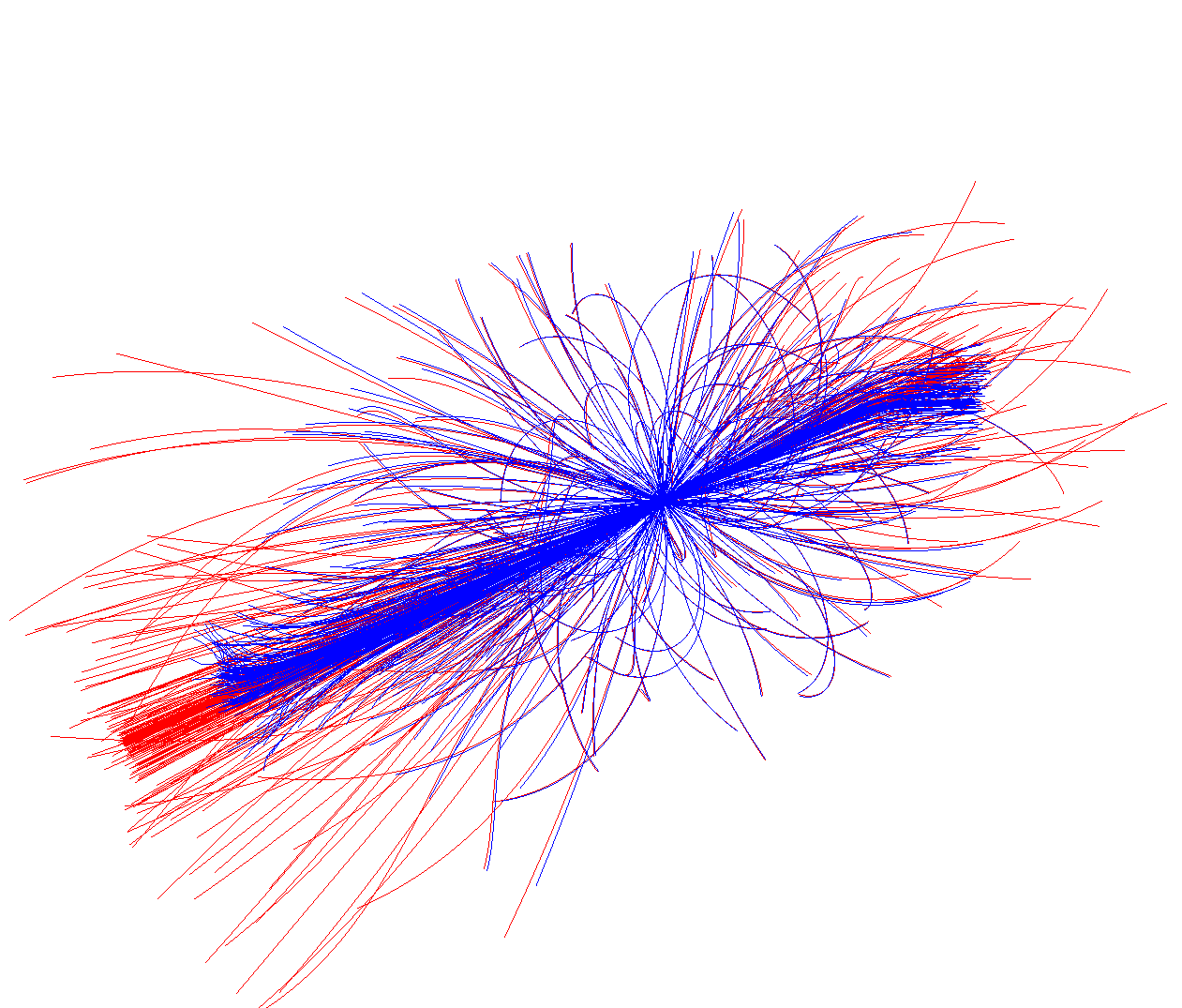}
    \caption{Sample rendering of 500 propagated tracks using constant (red) and accurate (blue) magnetic field models. Differences in track positions and curvatures are visible when the accurate model is used.}
    \label{fig:visual_inspection}
\end{figure}

To visually examine the results of our propagator, we have changed the render type from points to lines in our shader and enabled rasterization. Figure \ref{fig:visual_inspection} shows a sample of 500 particles propagated with the constant field (in red) and with the Chebyshev model (in blue) superimposed on top of each other. In the picture, a slight difference in curvature can be seen for helical tracks. By focusing on the muon tracks (straight lines going through the collision center) we can see the influence of the Muon Arm magnet, which bends them some distance away of the volume of the central barrel (on the left side in the picture). The constant field does not take into account the existence of this second magnet, so muon tracks in this case remain straight.

\begin{table}[ht]
    \centering
    \sisetup{scientific-notation=true, round-mode=places, round-precision=3, output-exponent-marker = \text{e}, table-format=1.3e+1}
    \begin{tabular}{ l S S S S }
    \toprule
    & \multicolumn{4}{c}{RMSE [\unit{\centi\metre}]}\\
    \cmidrule(lr){2-5}
    Algorithm & x & y & z & Total\\
    \midrule
    \texttt{const} & 0.00140716 & 0.00138117 & 0.00107026 & 0.001286197
\\
    \texttt{const\_barrel} & 0.00140716 & 0.00138117 & 0.00107026 & 0.001286197\\
    \texttt{ssbo} & 3.9571 & 4.1255 & 6.57067 & 4,884423333
\\
    \texttt{texture} & 3.95638 & 4.12416 & 6.56236 & 4,880966667
\\
    \texttt{sparse} & 3.95638 & 4.12416 & 6.56236 & 4,880966667\\
    \texttt{glsl} & 3.95567 & 4.13587 & 6.56937 & 4.88697
\\
    \bottomrule
    \end{tabular}
    \caption{Difference between original implementation (with constant field) and GPU implementation with various detailed field approaches (see Section~\ref{sec:impl}) as Root Mean Square Error for 7250 tracks.}
    \label{tab:rmse}
\end{table}

\section{Conclusion}
\label{sec:conc}

By studying the particle propagator algorithm from the ROOT framework we were able to implement a GLSL shader version, executable on the GPU. We verified the code correctness by fetching generated trajectories from the GPU via OpenGL \texttt{Transform Feedback} feature and compared them with trajectories generated by the original code --- obtained minuscule discrepancies were expected due to rounding errors of floating point arithmetic. We later replaced the constant field model used in the propagator with a detailed implementation studied in \cite{nowakowski}, which not only takes into account the volume of the central barrel magnet and slight imperfections of its generated field but models also a second magnet present in the Muon Arm, which is completely omitted in the constant field model. This advanced implementation offers a couple of approaches of model storage (buffer, texture, sparse texture, computation-based), which were all used in performance tests.

All advanced (i.e. non-constant) approaches performed similarly in the track deviation test (\SI{5}{\centi\meter} on average per track). The performance varied greatly depending both on the method as well as the graphics card used. If we consider the typical amount of tracks, then on the older card (GTX 1050 Ti):
\begin{itemize}
    \item constant methods achieved 250 frames per second,
    \item shader storage buffer method achieved 175 frames per second, 
    \item texture-based methods achieved 50 frames per second,
    \item Chebyshev method achieved 19 frames per second.
\end{itemize}
\noindent We consider every solution other than the last perfectly suitable for an interactive visualisation application. We deem the last result as acceptable, but to maximize user experience we suggest using one of the faster methods during the majority of time spent with such software and switching to the Chebyshev method (as it is the most accurate~\cite{nowakowski}) only for producing a final set of pictures for a screenshot or a video.

For the same amount of tracks, the results obtained on RTX 2080 were much closer to each other in terms of frame rates:
\begin{itemize}
    \item constant methods and shader storage buffer method achieved 208 frames per second,
    \item plain texture method achieved 208 frames per second as well,
    \item sparse texture method achieved 128 frames per second,
    \item Chebyshev method achieved 109 frames per second.
\end{itemize}
\noindent This GPU produced very good, real-time frame rates across the board. More than 60 frames per second regardless of the used method guarantees a smooth user experience in an interactive visualisation application.

In conclusion, the work done in this paper can fully replace the basic functionality of the Event Display. Performing the particle propagation on the GPU, besides parallelization of the work (with the geometry shader, each core propagates its own particle), enables tweaking both the propagator and magnetic field's parameters and watching the resulting changes in track shape in real-time. Because of this our software could be used not only by physicists, but also as a teaching aid for all levels of education, for example in the MasterClass projects~\cite{alicemasterclass,lhcbmasterclass,cmsmasterclass}. Although we have used the simplest possible output geometry in our study, the shader can be easily modified to show other parameters of the propagated particle on top of the tracks, such as velocity vector, electromagnetic force etc.

Lastly, although ALICE--focused, our work could be used to improve visualisation software used in other detector experiments, due to the ROOT framework being universally used across CERN~\cite{root7,rooteve}.

\section{Acknowledgements}
We would like to thank the ALICE Collaboration for guidance and support during our research as well as for the access to all software.

This work was supported by the Polish National Science Centre under agreements no. UMO-2016/21/D/ST6/01946, no. UMO-2022/45/B/ST2/02029, no. 2021/43/D/ST2/02214, by the Polish Ministry for Education and Science under agreements no. 2022/WK/01 and 5236/CERN/2022/0, as well as by the IDUB-POB-FWEiTE-1 project granted by Warsaw University of Technology under the program Excellence Initiative: Research University (ID-UB).

\bibliographystyle{unsrtnat}

\bibliography{cas-refs}


\end{document}